\documentclass[letterpaper]{jpconf}
\usepackage{graphicx}

\begin{document}

\title{Evaluation of the semiclassical coherent state propagator
in the presence of phase space caustics}

\author{A D Ribeiro$^{1,2}$ and M A M de Aguiar$^2$}

\address{$^1$ Instituto de F\'{\i}sica, Universidade de S\~ao
Paulo, Usp, 05315-970, S\~ao Paulo, S\~ao Paulo, Brazil}
\address{$^2$ Instituto de F\'{\i}sica ``Gleb Wataghin'',
Universidade Estadual de Campinas,Unicamp, 13083-970, Campinas,
S\~ao Paulo, Brazil}

\ead{aribeiro@ifi.unicamp.br,aguiar@ifi.unicamp.br}

\begin{abstract}
A uniform approximation for the coherent state propagator, valid in
the vicinity of phase space caustics, was recently obtained using
the Maslov method combined with a dual representation for coherent
states. In this paper we review the derivation of this formula and
apply it to two model systems: the one-dimensional quartic
oscillator and a two-dimensional chaotic system.
\end{abstract}

\section{Introduction}

The representation of coherent states has been used to describe a
wide variety of physical systems~\cite{klauderbook}. In particular,
coherent states provide a natural phase space representation of
quantum mechanics and are specially well suited to the study of the
semiclassical limit. The coherent state representation was first
formalized by Bargmann in 1961 \cite{bargmann} and later used by
Glauber \cite{glauber} to describe the electromagnetic field in
quantum electrodynamics. Fairly complete review articles on
coherent states and applications can be found in
references~\cite{klauderbook,perelomov,zhang}.

The quantum propagator $K(z''^*,z',T)\equiv\langle
z''|e^{-i\hat{H}T/\hbar}|z'\rangle$ represents the probability
amplitude that an initial coherent state $|z'\rangle$ evolves into
another coherent state $|z''\rangle$ after a time $T$. A path
integral formulation for this propagator was introduced by
Klauder~\cite{klauderpi}. The paths contributing to $K(z''^*,z',T)$
are those connecting $(q',p')\equiv(\langle z'|\hat{q}|z'\rangle,
\langle z' |\hat{p}|z'\rangle)$ to $(q'',p'')\equiv(\langle
z''|\hat{q}|z''\rangle, \langle z'' |\hat{p}|z''\rangle)$. In the
semiclassical limit, it turns out that the most important paths are
complex classical trajectories governed by the hamiltonian function
$\tilde{H}\equiv\langle z|\hat{H}|z\rangle$, with boundary
conditions involving the average values $q'$, $p'$, $q''$ and
$p''$. Klauder was the first to consider this type of
approximation~\cite{Klau79}, being followed by a number of other
contributors~\cite{Weis82b,solari, kochetov98}. More recently, a
detailed derivation of the semiclassical coherent state propagator
for one dimensional systems was published~\cite{Bar01}. In the last
two decades much numerical and analytical work has been done in
semiclassical methods with coherent states for
one~\cite{adachi,Klau95,xavier,gross,tanaka,f2003,marcelPRA,marcelJMP}
and two~\cite{vanv,ribeiro1} dimensional systems. A recent review
can be found in reference~\cite{dimitri}.

While the semiclassical propagator is usually very accurate for
short times, phase space caustics and non-contributing trajectories
inevitably appear as $T$ increases, introducing large errors and
imperfections~\cite{adachi,Klau95,tanaka,vanv,ribeiro1}.
Non-contributing trajectories must be identified and excluded from
the calculation because their contributions to the propagator are
non-physical. From the mathematical point of view, non-contributing
trajectories correspond to forbidden deformations of the contour of
integration necessary to carry out the stationary phase
approximation that leads to the semiclassical formula. Phase space
caustics, on the other hand, are special points where the amplitude
of the semiclassical propagator diverges and the approximation
simply breaks down. In order to calculate the propagator in the
vicinity of caustics one needs to improve the semiclassical
approximation, going beyond the usual quadratic expansion.

In two recent papers~\cite{prl,rib07a} we have derived a uniform
approximation for the coherent state propagator that remains finite
in the presence of phase space caustics. The derivation involved
the introduction of a dual representation for the coherent states
and the method of Maslov~\cite{maslov1}. In the present article, we
review the formalism used in these previous papers and apply it to
the one-dimensional quartic oscillator and to the two dimensional
chaotic Nelson potential. We show that the uniform formula
completely eliminates the divergences caused by the caustics,
providing a very accurate semiclassical description of the
propagator in these regions.

This paper is organized as follows: in the next section we briefly
review the representation of coherent states and the quantum
propagator. Section~\ref{secondorder} describes the semiclassical
approximation to the propagator based on a second order expansion
around stationary trajectories. The dual representation for
coherent states is introduced in section~\ref{dualrep} and used in
section~\ref{uniformapp} to derive the uniform approximation. In
sections~\ref{quart} and \ref{nelson} we present numerical
applications of the uniform formula and, in section~\ref{final}, we
present our final remarks.

\section{The coherent state propagator}
\label{csp}

Let $H_0=\hbar\omega(\hat{a}^{\dagger}\hat{a}+1/2)$ be the
Hamiltonian of a harmonic oscillator of mass $m$ and frequency
$\omega$. The normalized coherent states of $H_0$ are defined by
\cite{klauderbook, perelomov,zhang}
\begin{equation}
  \label{coh}
  |z \rangle = e^{-\frac{1}{2} |z|^2}
   e^{z \, \hat{a}^{\dagger}} |0 \rangle ,
\end{equation}
where
\begin{equation}
\hat{a}=\frac{1}{\sqrt{2}}
\left(\frac{\hat{q}}{b}+i\frac{\hat{p}}{c}\right) \;,
\quad \quad
z=\frac{1}{\sqrt{2}}
\left(\frac{q}{b}+i\frac{p}{c}\right)
\label{csr}
\end{equation}
and $|0\rangle$ is the oscillator's ground state. The real labels
$q$ and $p$ are the average values of the position and momentum
operators respectively and the length scales
$b=\sqrt{\hbar/(m\omega)}$ and $c=\sqrt{m\hbar\omega}$ satisfy
$b\,c=\hbar$. Three important properties of the coherent states are
(over)completeness, overlap relation and eigenvalue equation:
\begin{equation}
\label{one}
\mathbf{1}=\int|z\rangle\frac{d^2 z}{\pi}\langle z|=
\int|z\rangle\frac{dq\,dp}{2\pi\hbar}\langle z| ~,
\end{equation}
\begin{equation}
\displaystyle{
\langle z_i|z_j\rangle=
\exp{\{-\frac{1}{2}|z_i|^2+z_i^*z_j-\frac{1}{2}|z_j|^2}\}}
\label{ics}
\end{equation}
and
\begin{equation}
\label{eig}
\hat{a} |z\rangle = z |z\rangle.
\end{equation}

It will also be important in the derivation of our uniform
approximation to define non-normalized coherent states, or Bargmann
states \cite{bargmann}, by
\begin{equation}
  \label{barg}
  |z )= e^{z \, \hat{a}^{\dagger}} |0 \rangle.
\end{equation}
For these states the unit operator and the overlap equation become
\begin{equation}
\label{oneb}
\mathbf{1}=\int|z) \frac{e^{-|z|^2}}{\pi}
(z| \,d^2 z \qquad \mbox{ and} \qquad (z_i|z_j) = e^{z_i^*z_j}.
\end{equation}

The quantum propagator in the Bargmann and the coherent states
representations are given, respectively, by
\begin{equation}
k({z''}^*,z',T)= (z''|e^{-i\hat{H}T/\hbar}|z')
\end{equation}
and
\begin{equation}
K({z''}^*,z',T)=\langle z''|e^{-i\hat{H}T/\hbar}|z'\rangle =
e^{-\frac{1}{2}|z'|^2 - \frac{1}{2}|z''|^2} k({z''}^*,z',T).
\end{equation}
These two quantities contain the same physical information and
differ only in the normalization. For the purposes of the theory to
be developed in sections~\ref{dualrep} and \ref{uniformapp} it
shall be more convenient to work with the Bargmann states.

The propagator $K({z''}^*,z',T)$ can be written in terms of path
integrals, from which standard semiclassical approximations can be
performed. Here we present a very brief summary of path integral
formulation, referring to~\cite{Bar01} for the details. The first
step is to divide the propagation time $T$ into $N$ small intervals
of size $\epsilon = T/N$ so that
\begin{equation}
K({z''}^*,z',T)=\lim_{N\rightarrow\infty}
\langle z''|
\underbrace{e^{-i\hat{H}\epsilon/\hbar}\ldots e^{-i\hat{H}\epsilon/\hbar}}_{N\,\mathrm{times}}
|z'\rangle .
\end{equation}
Next, the coherent state unity operator~(\ref{ics}) is inserted
between each infinitesimal operator $e^{-i\hat{H}\epsilon/\hbar}$.
The path integral formula, obtained by evaluating the expression
for each resulting infinitesimal propagator $\langle
z_{j+1}|e^{-i\hat{H}\epsilon/\hbar}|z_j\rangle$, reads as
\begin{equation}
K({z''}^*,z',T)=\lim_{N\rightarrow\infty}
\int \Bigg\{ \prod_{j=1}^{N-1}\frac{d^2 z_j}{\pi}\Bigg\}
e^{\frac{i}{\hbar}\sum_{k=0}^{N-1}\epsilon
\left[
\frac{i\hbar}{2}
\left(\frac{z_{k+1}-z_k}{\epsilon}~z_{k+1}^{*}-\frac{z_{k+1}^{*}-z_{k}^{*}}{\epsilon}~z_{k}\right)
-  \tilde{H}_{k + \frac{1}{2}}
\right]},
\label{pi}
\end{equation}
where we have defined $\tilde{H}_{k+1/2}=\langle
z_{k+1}|\hat{H}|z_{k}\rangle/\langle z_{k+1}|z_{k}\rangle$ and
identified $|z'\rangle=|z_0\rangle$ and $\langle z''|=\langle
z_N|$. Equation~(\ref{pi}) represents the path integral formula of
the quantum propagator. Written as a function of the numbers
$(q_j,p_j)$, the propagator becomes an infinite sum of
contributions of all possible phase space paths linking the initial
point $(q',p')$ to the final $(q'',p'')$.

\section{Second order semiclassical approximation}
\label{secondorder}

By taking the formal semiclassical limit $\hbar\rightarrow0$, the
integrals~(\ref{pi}) can be performed~\cite{Bar01} by the saddle
point method~\cite{bleistein}. It can be shown that critical paths,
those whose contribution to the integral are relevant, are
classical trajectories $(Q(t),P(t))$, satisfying the boundary
conditions
\begin{equation}
\frac{Q(0)}{b}+i\frac{P(0)}{c}=\frac{q'}{b}+i\frac{p'}{c}
\quad\mathrm{and}\quad
\frac{Q(T)}{b}-i\frac{P(T)}{c}=\frac{q''}{b}-i\frac{p''}{c},
\end{equation}
and governed by the average hamiltonian
$\tilde H=\langle z|\hat H|z\rangle$. One might think initially
that the trajectory starting at $(Q(0),P(0))=(q',p')$ and ending at
$(Q(T),P(T))=(q'',p'')$ would be the only solution to these
equations. However, these boundary conditions are very restrictive
and such a trajectory usually does not exist: indeed, giving the
{\em initial position and initial momentum}, the trajectory is
completely determined so that the final point $(Q(T),P(T))$ is
generally different from $(q'',p'')$. This means that, in general,
there is no {\em real} critical path to the integral~(\ref{pi}).
Complex trajectories, however, can usually be found if we
analytically extend the integration to the complex phase space,
letting $Q$ and $P$ be complex variables.

In this case it is more convenient to introduce new variables $u$
and $v$, instead of using the complex position $Q$ and momentum
$P$, such that
\begin{equation}
u=\frac{1}{\sqrt{2}}\left(\frac{Q}{b}+i\frac{P}{c}\right)
\quad \mathrm{and} \quad
v=\frac{1}{\sqrt{2}}\left(\frac{Q}{b}-i\frac{P}{c}\right) .
\label{nv}
\end{equation}
In terms of $u$ and $v$ the classical equations of motion become
\begin{equation}
\dot u= \frac{1}{i\hbar}\frac{\partial\tilde H}{\partial v}
\quad \mathrm{and} \quad
\dot v= -\frac{1}{i\hbar}\frac{\partial\tilde H}{\partial u},
\label{emuv}
\end{equation}
where $\tilde H = \langle v |\hat{H}|u\rangle$, and the boundary
conditions assume a simpler form,
\begin{equation}
\begin{array}{l}
\displaystyle u(0)=\frac{1}{\sqrt{2}}\left(\frac{Q(0)}{b}+i\frac{P(0)}{c}\right)
=\frac{1}{\sqrt{2}}\left(\frac{q'}{b}+i\frac{p'}{c}\right)=z',\\
\displaystyle v(T)=\frac{1}{\sqrt{2}}\left(\frac{Q(T)}{b}-i\frac{P(T)}{c}\right)
=\frac{1}{\sqrt{2}}\left(\frac{q''}{b}-i\frac{p''}{c}\right)=z''^*.
\end{array}
\label{bbuv}
\end{equation}
Notice that this {\em does not} imply that $v(0)=z'^*$ and
$u(T)=z''$. Given $u(0)=z'$ and $v(T)={z''}^*$ a complex trajectory
$(u(t),v(t))$ can be calculated and the values of $v(0)$ and
$u(T)$ come out of this calculation. In general there might be more
than one trajectory governed by~(\ref{emuv}) and
satisfying~(\ref{bbuv}).

Returning to Eq.~(\ref{pi}), by expanding the exponent up to second
order around the complex classical path, we find the following
semiclassical formula for the propagator
\begin{equation}
K_{sc}^{(2)}({z''}^*,z',T)=
\sum_{\mathrm{traj.}}
\sqrt{\frac{1}{M_{vv}}}~
\exp{\left\{\frac{i}{\hbar}~(\mathcal{S}+\mathcal{G}) -\frac{1}{2}\left(|z'|^2+|z''|^2\right)\right\}},
\label{sp}
\end{equation}
where the index ${(2)}$ means ``second order expansion''. The sum
in Eq.~(\ref{sp}) is over the complex classical trajectories, as
discussed, and
\begin{eqnarray}
\mathcal{S}({z''}^*,z', T)&=&\int_{0}^{T}
\left[\frac{i\hbar}{2}\left(\dot{u}~v - u~\dot{v}\right)-\tilde{H}\right] dt
-\frac{i\hbar}{2}\left[{  u(T) {z''}^* + z'v(0) } \right],\label{action}\\
\mathcal{G}({z''}^*,z', T) &=&\frac{1}{2} \int_{0}^{T}
\frac{\partial^{2}\tilde{H}}{\partial u \; \partial v} \, dt .
\label{gfactor}
\end{eqnarray}
Finally $M_{vv}$ is an element of the tangent matrix $M$ defined by
\begin{eqnarray}
\left(\begin{array}{c}\delta{u}(T)\\\delta{v}(T)\\\end{array}\right) =
\left(\begin{array}{cc}
M_{uu}&M_{uv}\\M_{vu}&M_{vv}\\
\end{array}\right)
\left(\begin{array}{c}\delta {u}(0)\\\delta {v}(0)\\\end{array}\right) \, ,
\label{mmatrix}
\end{eqnarray}
where $\delta {u}$ and $\delta {v}$ are small displacements around
the classical trajectory. The elements of the tangent matrix are
related to second derivatives of the action $\mathcal
S$~\cite{Bar01}. The phase of $M_{vv}$ contains important Maslov
phases.

The corresponding semiclassical formula for $k({z''}^*,z',T)$ is
identical, except that it does not have the normalization factor in
the exponent.

In the strict limit where $\hbar\rightarrow 0$, the semiclassical
propagator becomes a delta function at the phase space point
$(q'',p'')$ linked to $(q',p')$ by a real classical trajectory.
Therefore, for small but finite $\hbar$, we expect large
contributions to the propagator arising from nearly real
trajectories. The more the trajectory wanders into the complex $p$
and $q$ space, the less it should contribute to
$K_{sc}^{(2)}({z''}^*,z',T)$. By inspection of Eq.~(\ref{sp}), we
see that this statement is true only if the real part of the
exponent in (\ref{sp}) is negative. Trajectories for which such
real part is positive furnish non-physical contributions to the
propagator that become arbitrarily large when $\hbar$ goes to zero.
These are non-contributing trajectories and must be excluded from
the calculation. They correspond to complex critical paths whose
steepest descent contour of integration cannot be reached by
deformations allowed by Cauchy's theorem. Their structure are
closely related to the {\it Stokes
Phenomenon}~\cite{stokes,phenomenon}, discussed in a number of
papers on semiclassical
approximations~\cite{adachi,Klau95,vanv,ribeiro1,shudo,Agu05,parisio2}.
In particular, Ref.~\cite{parisio2} shows an explicit example where
non-contributing trajectories arise from forbidden deformations of
the original contour of integration.

Another common source of complications in semiclassical formulas
are focal points or caustics. These are special points where the
second order approximation breaks down because of singularities in
the formula's pre-factor. In the case of
$K_{sc}^{(2)}({z''}^*,z',T)$ this happens when $M_{vv}=0$.
According to Eq.~(\ref{mmatrix}), if $M_{vv}$ goes to 0, we can set
small initial displacements $\delta u(0)=0$ and $\delta v(0)\neq 0$
such that $\delta u(T)\neq 0$ and $\delta v(T)=0$, implying that
there are at least two nearby trajectories satisfying the correct
boundary conditions~(\ref{bbuv}). The point where these
trajectories coalesce is called a {\it phase space caustic} and the
semiclassical formula~(\ref{sp}) fails in its vicinity. Contrary to
non-contributing trajectories, trajectories going through caustics
cannot simply be excluded, since the problem lies on the approach
used, and not on the orbit itself. Thus, to evaluate the
semiclassical propagator in the vicinity of phase space caustics
one needs better approximations, beyond the second order. We derive
such an approximation in the next two sections.

\section{Dual representation for the coherent state propagator}
\label{dualrep}

The most direct way to improve the quadratic approximation is to go
back to Eq.~(\ref{pi}) and expand the exponent of the propagator to
third order around the critical paths. This, however, would be
extremely complicated, since the integral~(\ref{pi}) is
multi-dimensional. A simpler solution is to follow the method
proposed by Maslov~\cite{maslov1}. To illustrate the idea, suppose
that a semiclassical approximation for $\psi(q)$ has a singularity
at $q=q_0$. If we know the corresponding semiclassical formula for
$\psi$ in the momentum representation we can write
\begin{equation}
\psi_{sc}(q_0) = \int \langle q_0|p\rangle \psi_{sc}(p) dp =
\int \psi_{sc}(p) e^{ipq_0/\hbar} dp.
\label{masqp}
\end{equation}
If the integral over $p$ is performed by the usual second order
stationary phase approximation, the singularity in $\psi_{sc}(q_0)$
is recovered. However, doing a third order stationary phase
approximation produces a more accurate expression involving an Airy
function which remains finite at $q_0$~\cite{berry}. Therefore,
Maslov's method consists in finding the desired semiclassical
approximation in a conjugate representation and transforming back
to the original one by a third order expansion.

The problem in applying this idea to coherent states is that they
are defined in phase space and do not have a natural dual
representation. Since $z$ and $z^*$ play the role of conjugate
variables, we can think of $z$ as $q$ and $z^*$ as $p$, and the
coherent state propagator is always written in the mixed $p-q$
representation. It is impossible to write it in $q-q$ or $p-p$ forms,
since one cannot write a matrix element with two kets or two bras.
To overcome this difficulty we need to define a proper linear
application to play the role of the dual representation and we
shall do that using the non-normalized Bargmann representation.

Given the propagator $k({z''}^*,z',T)$, the associated dual
propagator is defined as
\begin{eqnarray}
\tilde k ( z ,z',T )=
\frac{1}{\sqrt{2 \pi i}}\int_{\tilde{C}} k({z''}^*,z',T) e^{-{z''}^*z} d{z''}^* ,
\label{ktil}
\end{eqnarray}
and the inverse application by
\begin{eqnarray}
k({z''}^*,z',T)=
\frac{1}{\sqrt{2 \pi i}}\int_{{C}} \tilde k(z,z',T) e^{{z''}^*z} dz,
\label{ktilinv}
\end{eqnarray}
where $\tilde{C}$ and $C$ are convenient paths as specified in
Ref.~\cite{prl}. These equations are reminiscent of (\ref{masqp}).
Notice that equation (\ref{ktil}) can also be written as
\begin{displaymath}
\tilde k ( z ,z',T )= \frac{1}{\sqrt{2 \pi i}} \int_{\tilde{C}}
\frac{(z''|e^{-i\hat{H}T/\hbar}| z')}{(z''|z)} d{z''}^* ,
\end{displaymath}
which can be interpreted as an attempt to `cancel' the bra $(z''|$
and introduce another ket $|z)$. Of course $\tilde{k}$ is not a
matrix element and, therefore, the above application is not a true
representation.

Equation~(\ref{ktilinv}) is the starting point to make improvements
on $k_{sc}^{(2)}(z''^*,z',T)$. In the regions where both propagators
are free of caustics, the semiclassical version of $\tilde
k(z,z',T)$ can be obtained by performing the integral~(\ref{ktil})
using the standard second order saddle point method with
$k(z''^*,z',T)$ replaced by $k^{(2)}_{sc}(z''^*,z',T)$. The result
is~\cite{prl}
\begin{eqnarray}
\tilde{k}_{sc}^{(2)}(z, z', T ) = \sum_{\mathrm{traj.}}
\sqrt{\frac{1}{M _{u v}}}~
\exp\left\{\frac{i}{\hbar} \tilde{\mathcal S}(z, z',T )
+ \frac{i}{\hbar} \tilde{\mathcal G}(z, z',T )
\right\}.
\label{ktil2}
\end{eqnarray}
The trajectories summed in this equation are not the same as those
of Eq.~(\ref{sp}), since they satisfy the conditions $u(0)=z'$ and
$u(T)=z$. As usual, $v(0)$ and $v(T)$ are not fixed, but come out
of the integration of Hamilton's equations with the above boundary
conditions. The tangent matrix element $M _{u v}$ is given by
Eq.~(\ref{mmatrix}) and $\tilde{\mathcal G}(z, z',T) $ is the
function $\mathcal{G}$ calculated with the new trajectory. The new
action $\tilde{\mathcal S}(z, z',T )=\mathcal S({z''}^*, z',T ) + i
\hbar z {z''}^*$ is the Legendre transform of $\mathcal S$, where
${z''}^* = {z''}^*(z,z',T)$ is obtained from the relation $-i\hbar
z =
\partial \mathcal{S}/\partial {z''}^*$.

According to Eq.~(\ref{mmatrix}), when $M _{v v}$ is zero, $M _{u
v}$ is usually not zero, which implies that if
${k}_{sc}^{(2)}({z''}^*, z', T )$ has a caustic for a complex
trajectory satisfying $u(0)=z'$, $v(T)={z''}^*$, implying values of
$u(T)$ and $v(0)$, then $\tilde{k}_{sc}^{(2)}(z, z', T )$ will not
have a caustic when calculated at the same trajectory, i.e., for
$u(0)=z'$ and $z=u(T)$.

Inserting~(\ref{ktil2}) into~(\ref{ktilinv}) leads to
\begin{eqnarray}
k_{sc} ({z''}^*,z',T )=
\frac{1}{\sqrt{2 \pi i} }\int_{\tilde{C}}
\sqrt{\frac{1}{M _{u v}}}{~e^{\frac{i}{\hbar}\tilde{\mathcal S}(z,z',T) +
\frac{i}{\hbar}\tilde{\mathcal G}+z{z''}^*
(z,z',T)}~dz},
\label{ktilinvsc}
\end{eqnarray}
where we omit the sum for simplicity. Eq.~(\ref{ktilinvsc}) is an
integral representation for the semiclassical coherent state
propagator. In order to calculate it we need to sum the
contributions of all trajectories starting at $u(0)=z'$ and ending
at $u(T)=z$ lying on the curve $\tilde{C}$. Eq.~(\ref{sp}) is
recovered if the second order saddle point method is applied
to~(\ref{ktilinvsc}), the critical paths being precisely the orbits
given by Eqs.~(\ref{emuv}) and (\ref{bbuv}).

Expanding the exponent of Eq.~(\ref{ktilinvsc}) to third order
around the critical paths leads to the so called regular
approximation derived in~\cite{rib07a}. The regular formula
provides satisfactory results only if the critical trajectories are
not too close to caustics, so that each trajectory still
contributes independently to the propagator. For the transitional
approximation, the exponent of (\ref{ktilinvsc}) is expanded around
the trajectory that lies exactly at the phase space caustic. Since
this trajectory is not a critical one, the formula is good only if
critical trajectories are sufficiently close to the caustic. In
this paper we will be concerned only with uniform approximations,
which provide a global semiclassical formula, reasonably accurate
over all the space spanned by the parameters $z'$, ${z''}^*$ and
$T$.

\section{Uniform Approximation}
\label{uniformapp}

A uniform approximation for the coherent state propagator was
obtained in Ref.~\cite{prl} for one-dimensional systems and in
Ref.~\cite{rib07a} for two dimensions. The expression presented
here is slightly different from those in Refs.~\cite{prl,rib07a},
as we point out below.

The simplest type of singularity that can appear in the
semiclassical propagator occurs when two nearby trajectories
coalesce at the caustic. In this case the function
$\phi(z)=i(\tilde\mathcal{S}(z,z',T) -i\hbar z {z''}^*)/\hbar$ has
two stationary points (corresponding to two complex trajectories
satisfying the same boundary conditions) that coalesce as $z''^*$
(here considered as a parameter) approaches the caustic. The basic
idea of the uniform approximation is to map this complicated
function into a simpler function with the same critical points and
the same behavior in the neighborhoods of these
points~\cite{uniform}. Since all that matters
in the semiclassical limit is the neighborhood of the critical
points, the integral with the new function should give about the
same results as the integral with the original function.

In the case of two coalescing trajectories the appropriate function
is $N(x)=\mathcal A-\mathcal Bx+\frac{1}{3}x^3$, which has two
saddle points at $\pm\sqrt{\mathcal B}$ that coalesce as $\mathcal B
\rightarrow 0$. Therefore we write [see Eq.~(\ref{ktilinvsc})]
\begin{eqnarray}
\int{\mathcal P(z,z',T)
e^{\frac{i}{\hbar}\tilde{\mathcal S}(z,z',T)+z{z''}^*}~dz}
= \int{~J(x) e^{N(x)}~dx},
\label{iu}
\end{eqnarray}
where the function $J(x)$ includes the Jacobian of the
transformation $z \rightarrow x$ and the contribution of the smooth
term $\mathcal P(z,z',T)=M _{u v}^{-1/2}
e^{\frac{i}{\hbar}\tilde{\mathcal G}(z,z',T)}$. In
Refs.~\cite{prl,rib07a} the logarithm of this term was also
included in the function $\phi(z)$. However, since $\mathcal P$
varies slowly with $\hbar$ it is reasonable to leave it out. From
the numerical point of view it turns out that the present
prescription is also more accurate than the ones
in~\cite{prl,rib07a}.

Imposing that the value of $N(x)$ at the saddle points
$\pm\sqrt{\mathcal B}$ coincide with the value of $\phi$ at
critical points $z_{1,2}$ (related to the two critical trajectories
of~(\ref{ktilinvsc})) we find
\begin{equation}
\mathcal A=\frac{i}{2\hbar}(\mathcal S_1+\mathcal S_2)
\quad \mathrm{and} \quad
\mathcal B=\left[\frac{3i}{4\hbar}(\mathcal S_2-\mathcal S_1)\right]^{2/3}, \label{ab}
\end{equation}
where $\mathcal S_i$ is the complex action of the trajectory
related to $z_{i}$. This determines completely the function $N(x)$.

Next we impose the equivalence between $N(x)$, in the vicinity of
$x=\pm\sqrt{\mathcal B}$, and $\phi(z)$, in the vicinity of
$z=z_{1,2}$. This is done by writing
\begin{equation}
\left.\left(\frac{\partial^2 N}{\partial x^2}\right)\right|_{\pm\sqrt{\mathcal B}}\delta x^2
= \pm2\sqrt{\mathcal B}~\delta x^2 =
\left.\left(\frac{\partial^2 \tilde{\mathcal{S}}}{\partial z^2}\right)\right|_{z_{1,2}}\delta z^2 ,
\end{equation}
which, by identifying~\cite{rib07a}
\begin{equation}
\left.\frac{\partial^2 \tilde\mathcal{S}}{\partial z^2}\right|_{z_{1,2}}
=i\hbar\left.\left(\frac{M_{vv}}{M_{uv}}\right)\right|_{z_{1,2}},
\end{equation}
implies that the jacobian of the transformation calculated at each
saddle point, $j_{1,2}$, amounts to
\begin{equation}
j_{1,2}=
\sqrt{\frac{\pm 2\sqrt{\mathcal B}}
{i\hbar\left.\left(M_{vv}/M_{uv}\right)\right|_{z_{1,2}}}}.
\end{equation}
The function $J(x)$, therefore, can be conveniently written as
\begin{equation}
J(x)=\frac{1}{2} \left( J_1 + J_2\right) -
\frac{x}{2\sqrt{\mathcal B}}\left( J_2 - J_1\right) + {\cal O}(x^2),
\end{equation}
where $J_i = \left.\left[M _{u v}^{-1/2} e^{\frac{i}{\hbar}{\mathcal G}
\left( {z''}^*, \,  z' ,T\right)}\right]\right|_{z_i} j_i$, so that when $x\rightarrow
x_{1,2} = \pm\sqrt{\mathcal{B}}$, we have $J(x_{1,2}) = J_{1,2}$.

Finally, we write the uniform formula for the propagator as
\begin{equation}
k_{sc}^{un} ({z''}^*, z',T)=
\frac{1}{\sqrt{2\pi i}}\int J(x) ~e^{\mathcal A- \mathcal
Bx+x^3/3}dx, \label{equn}
\end{equation}
or, discarding global phases,
\begin{eqnarray}
k_{sc}^{un} (z''^*, z',T)=
\sqrt\pi~e^{\mathcal A}
\left\{
\left(\frac{g_2-g_1}{\sqrt \mathcal B}\right)\mathrm{F'_i}(\mathcal B)+
(g_1+g_2)\mathrm{F_i}(\mathcal B)
\right\}, \label{uform}
\end{eqnarray}
where $g_{1,2}=\sqrt{\mp\sqrt{\mathcal B}/ \left.\left(
M_{vv}\right)\right|_{z_{1,2}}}~e^{\frac{i}{\hbar}{\left.\mathcal
G\right|_{z_{1,2}}}}$, $\mathrm{F'_i}$ is the derivative of
$\mathrm{F_i}$ with respect to its argument and $\mathrm{F_i}$ is
given by~\cite{bleistein}
\begin{equation}
\mathrm{F_{i}}(W)
=
\frac{1}{2\pi}\int_{C_i} \mathrm{d} t
\exp{ \left\{
i \left[ Wt  + \frac{1}{3} t^3 \right] \right\}} ,
\label{int7}
\end{equation}
for $\mathrm i=1,2,3$. The index $\mathrm i$ refers to three
possible paths of integration $C_i$, giving rise to three different
Airy functions~\cite{bleistein}. The correct path is determined
by Cauchy's theorem, but, in practice, we can use physical
criteria to justify the choice of $C_i$. The normalized propagator
$K_{sc}^{un} ({z}''^*, z',T)$ is obtained multiplying $k_{sc}^{un}$
by $e^{-\frac{1}{2}|z'|^2 - \frac{1}{2}|z''|^2}$.

To finish this section we emphasize that the application of the
uniform semiclassical formula (\ref{uform}) always involves the two
complex trajectories coalescing at the caustic and the choice of a
contour of integration. In the next two sections we present
numerical examples where these trajectories and the integration
path will be pointed out explicitly.

\section{The quartic oscillator}
\label{quart}

As a first application of the uniform approximation we consider the
one-dimensional quartic oscillator. This simple system illustrates
well the problems of non-contributing trajectories and phase space
caustics. Moreover, the structure of the complex phase space is
easy to visualize and shows clearly the complications that arise as
the propagation time increases. The hamiltonian is
\begin{equation}
\hat H = \frac{1}{2} \hat{P}^2 + \frac{1}{2}\hat{Q}^2 +B\hat{Q}^4,
\end{equation}
and we set $B=0.1$ and $\hbar=1.0$. The initial state $|z'\rangle$
is chosen at $q'=0$, $p'=-2.0$ with $b=1.0$, while the final
$|z''\rangle$ is given by $q''=0.5$, $p''=0.5$ with the same width
$b=1.0$. The calculation of $K_{sc}^{(2)}({z''}^*,z',T)$ for these
fixed states as function of $T$ shows that, for $T \approx 2.5$,
the semiclassical result has an unphysical peak, revealing the
presence of a caustic. We will show that this inaccurate result can
be controlled with the uniform formula~(\ref{uform}).

\begin{figure}
\centerline{
\includegraphics[width=16pc]{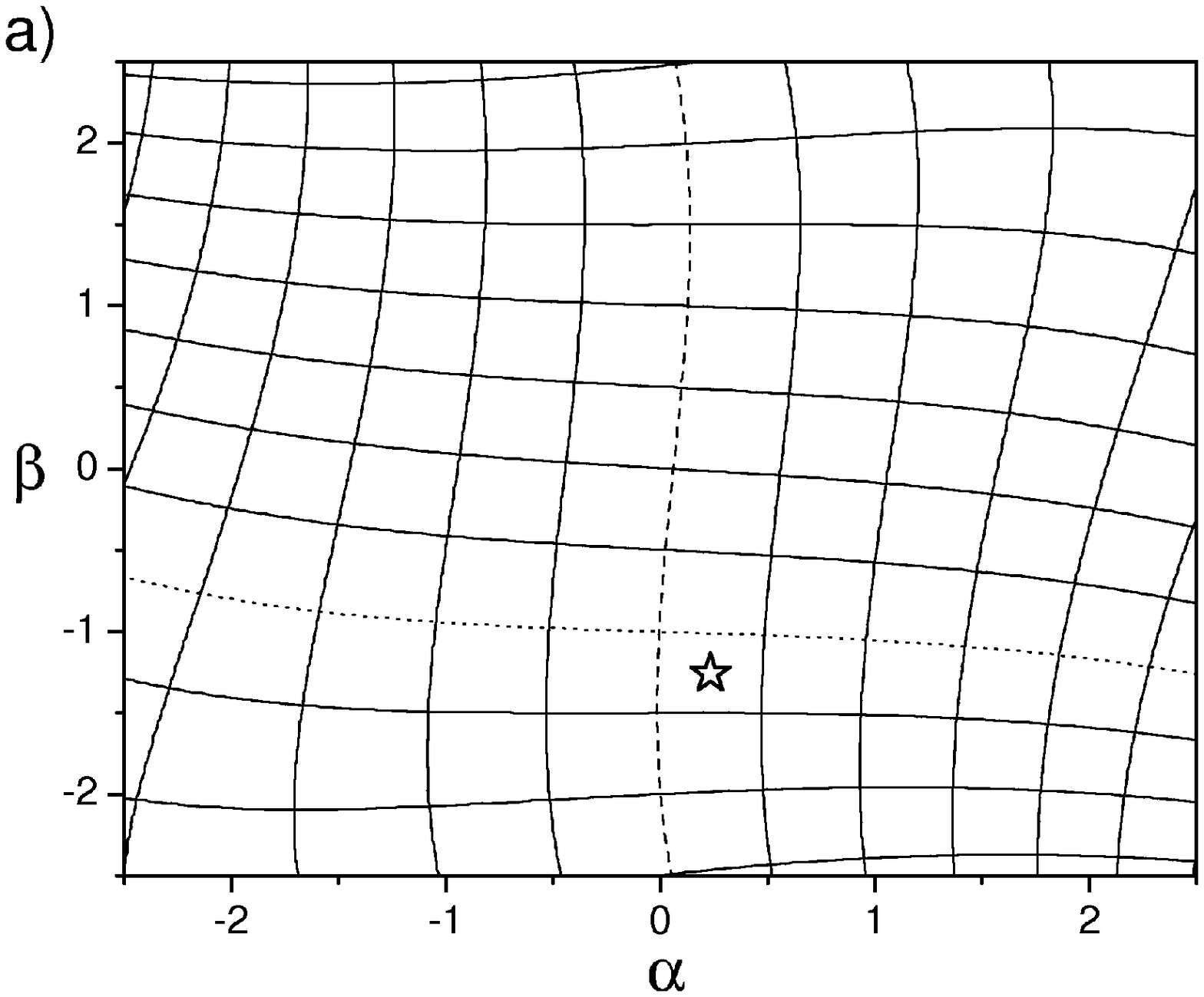}
\hspace{1cm}\includegraphics[width=16pc]{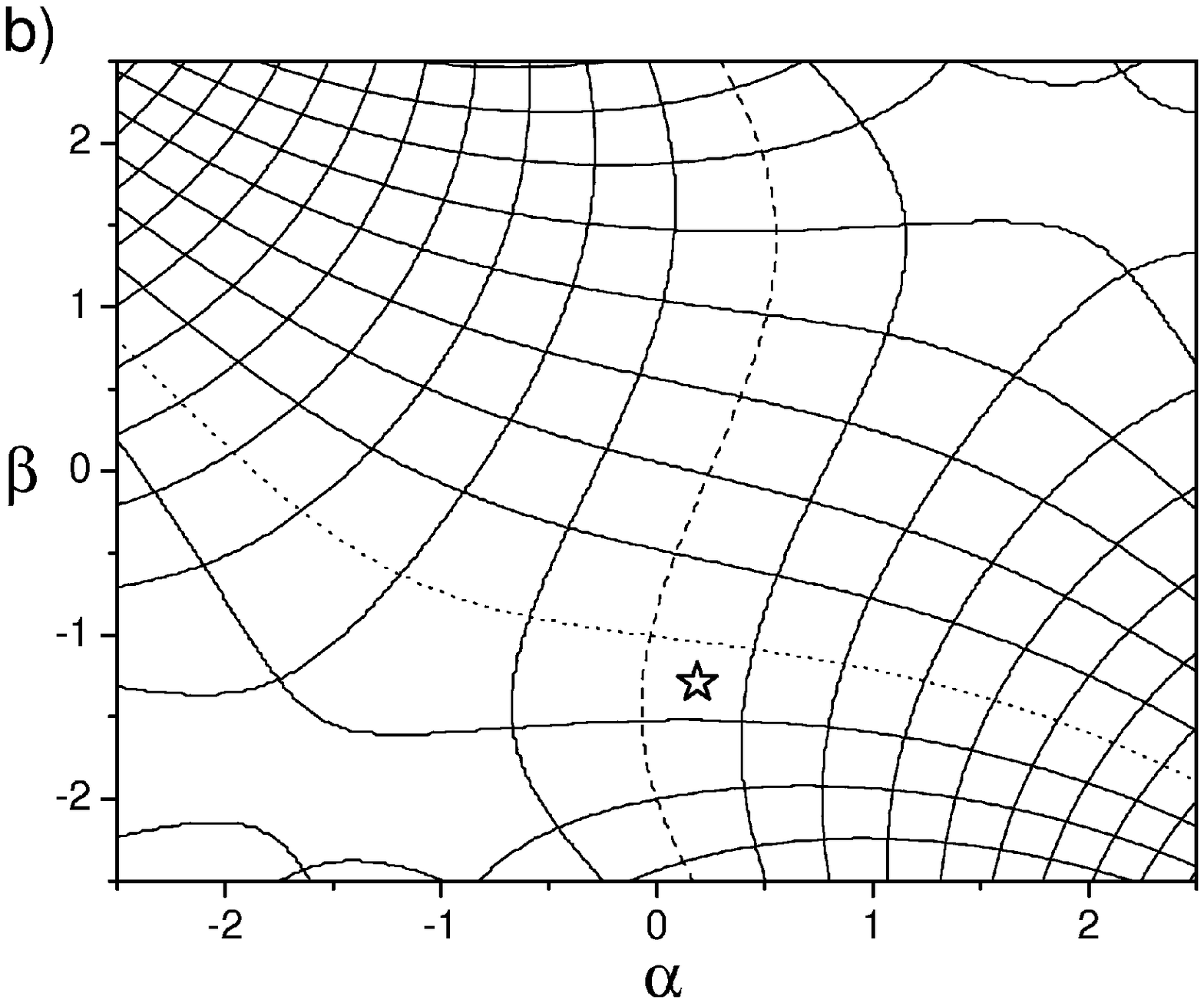}}
\vspace{.2cm}
\centerline{\includegraphics[width=16pc]{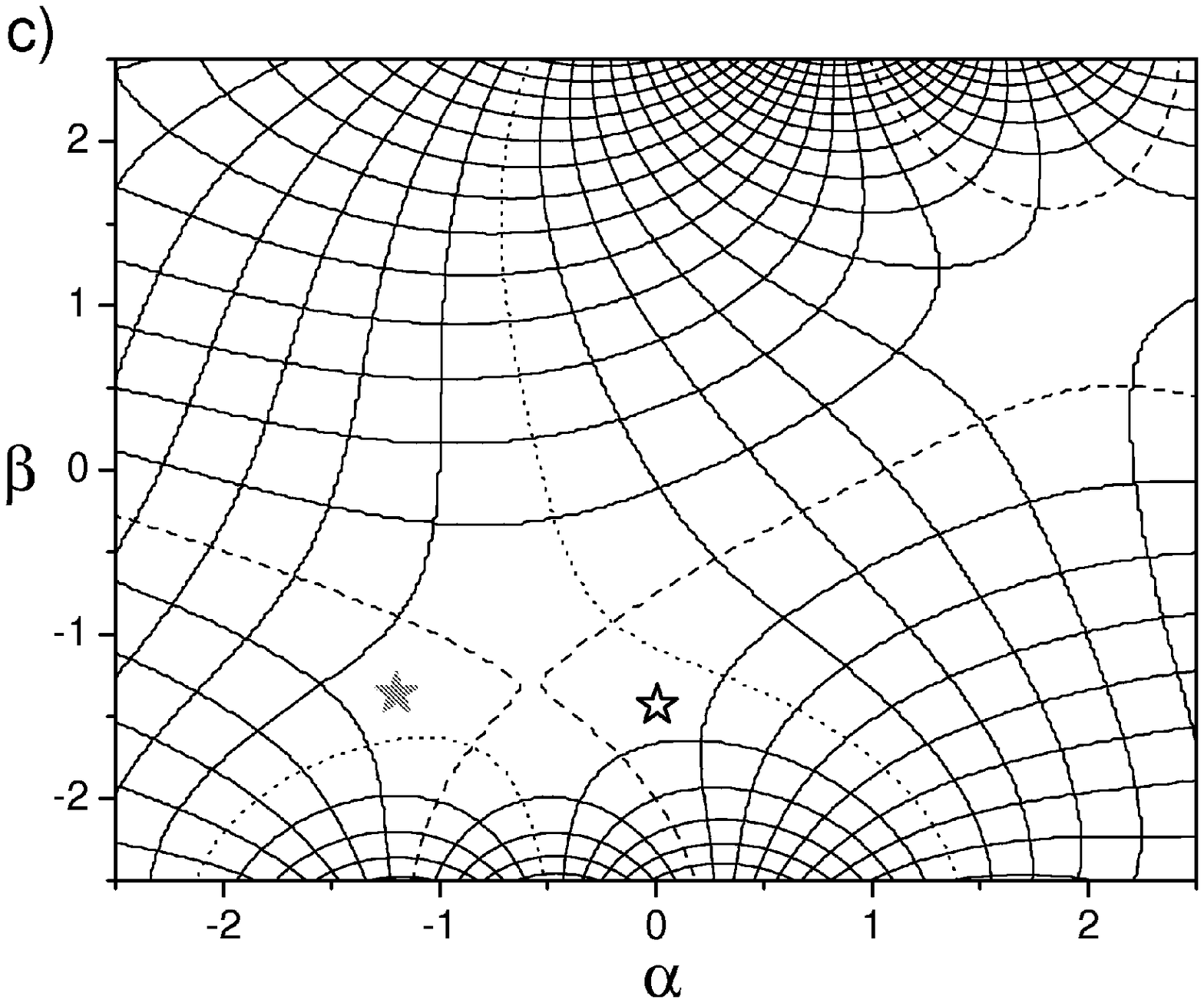}
\hspace{1cm}\includegraphics[width=16pc]{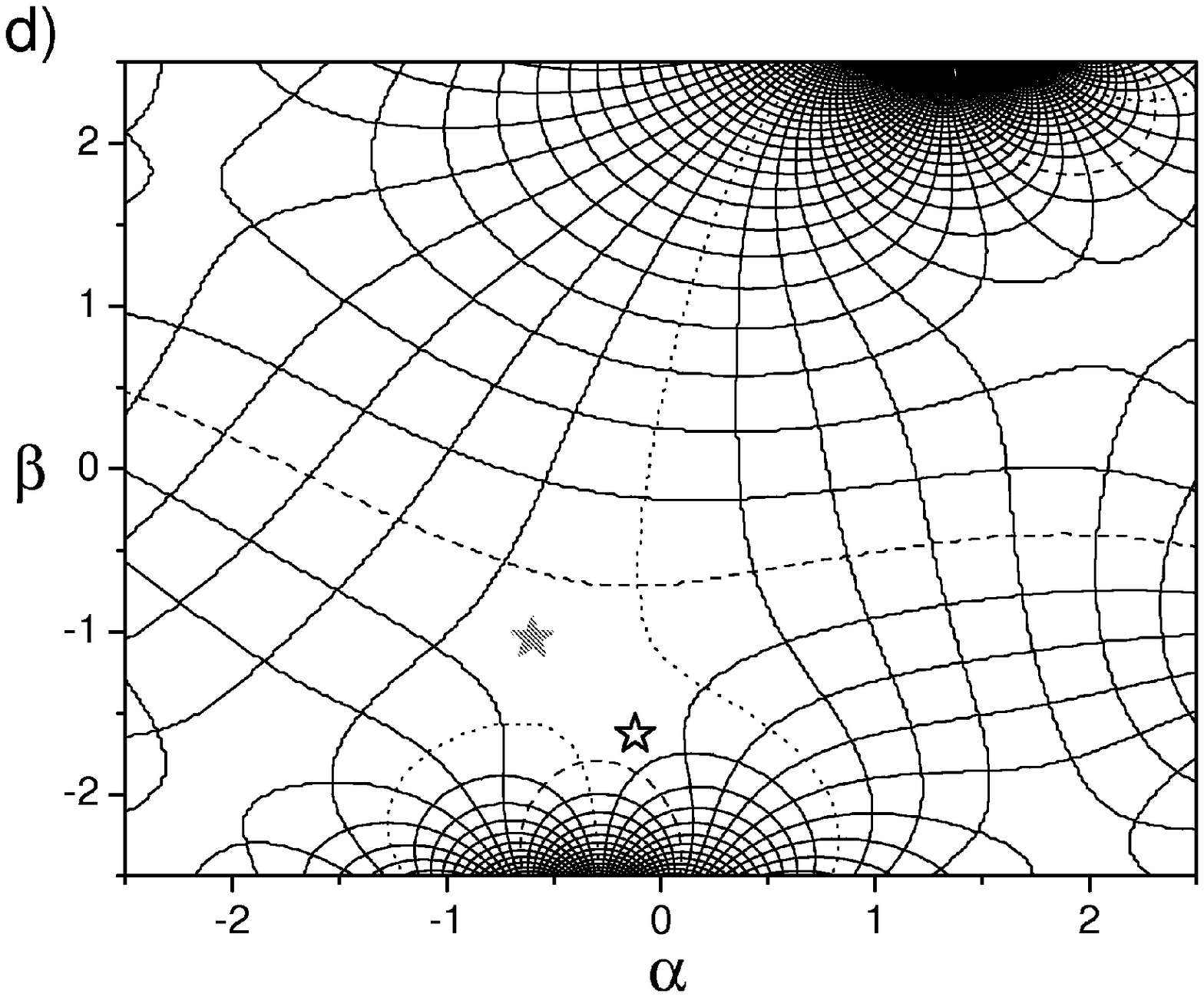}}
\vspace{.2cm}
\centerline{\includegraphics[width=16pc]{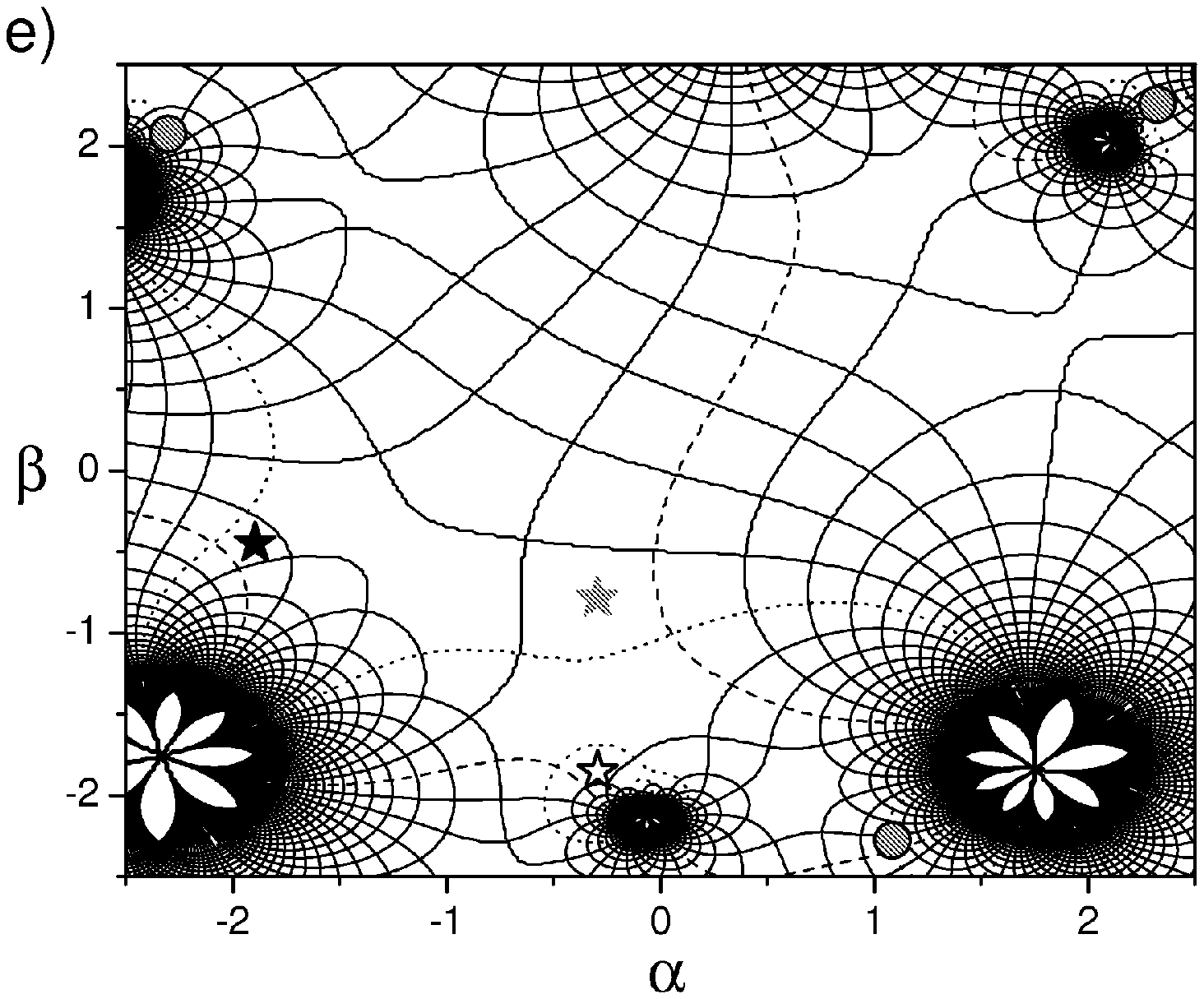}
\hspace{1cm}\includegraphics[width=16pc]{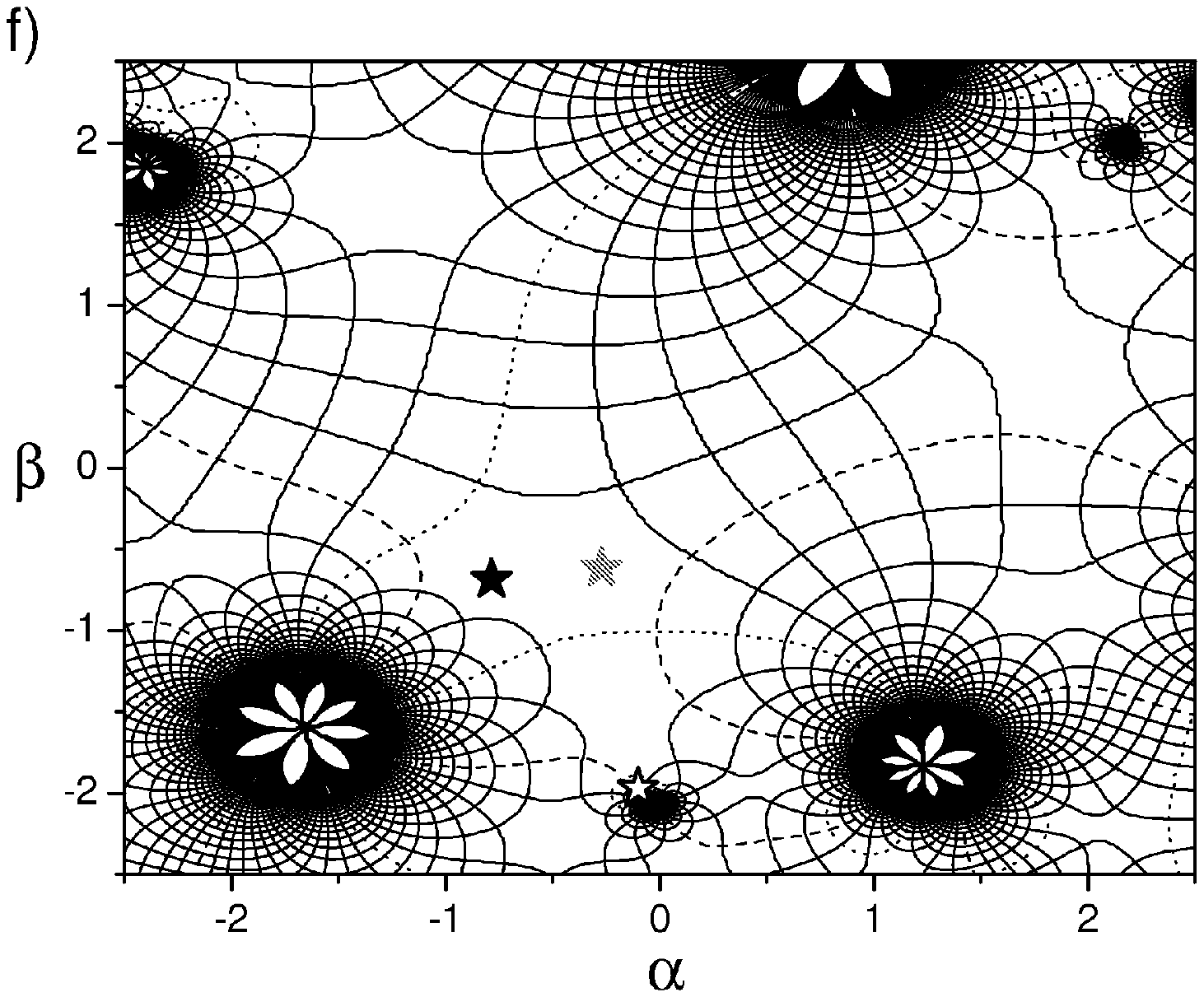}}
\caption{\label{fig1}
Curves of constant $Q''$ superimposed with those of constant $P''$ in the $w$ plane.
Dashed lines refer to $Q''=0$ and the dotted lines to $P''=0$. From panel (a) to
(f), the value of $T$ is to $T=0.06,\,0.24,\,0.70,\,1.02,\,2.20$ and $2.70$,
respectively. Stars are centered at the point $Q''=P''=0.5$. White stars represent
trajectories belonging to the family $f1$, grey stars to $f2$, and the black stars to $f3$.
The circles in panel (e) also represent potential candidates to be included in the sum~(\ref{sp}),
illustrating the existence of other trajectories satisfying the proper boundary conditions~(\ref{bbuv}).
In this case, since they lie far from the origin, they were not included in the calculation.}
\end{figure}

\begin{figure}
\centerline{\includegraphics[width=16pc]{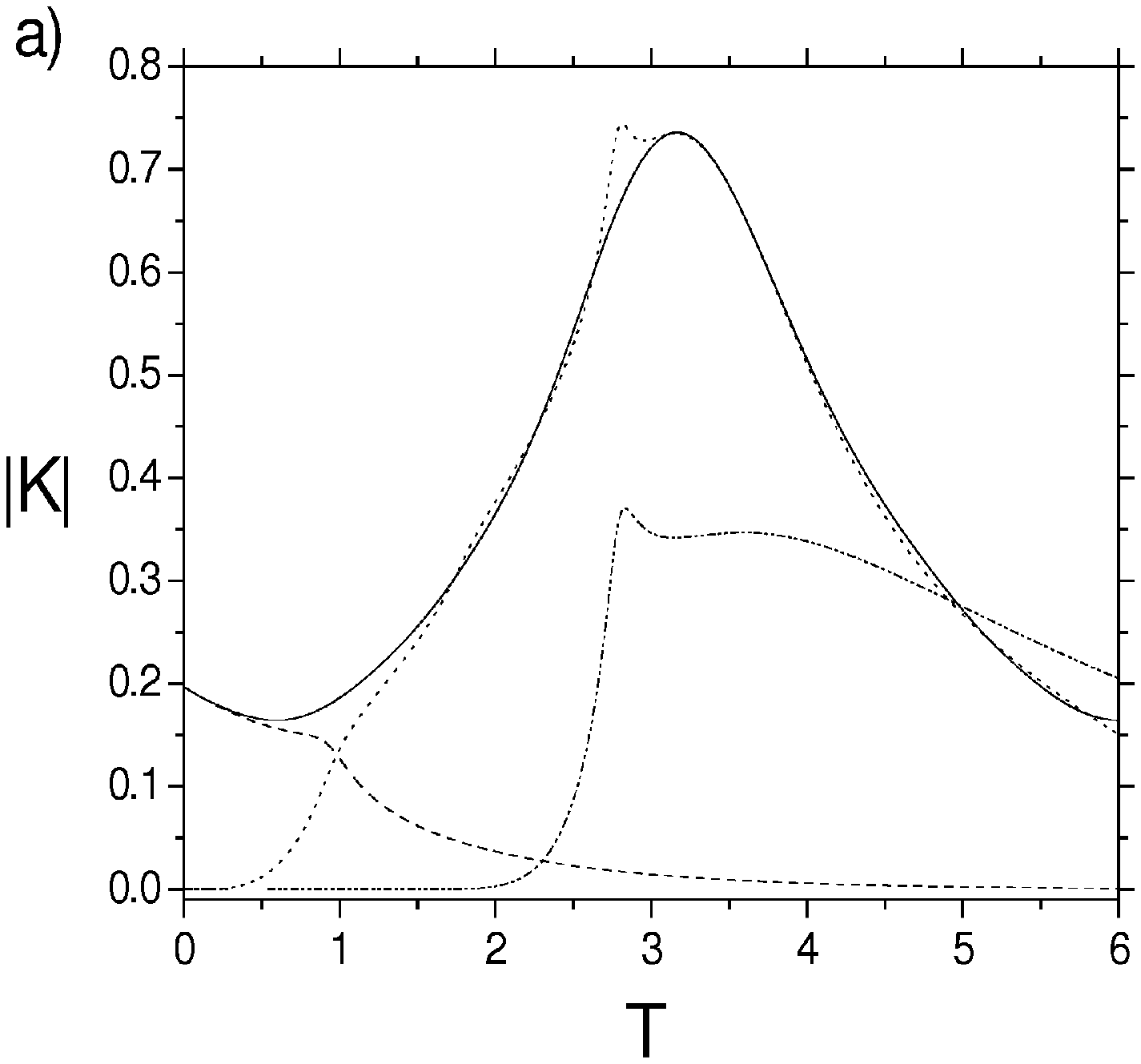}
\hspace{1cm}\includegraphics[width=16pc]{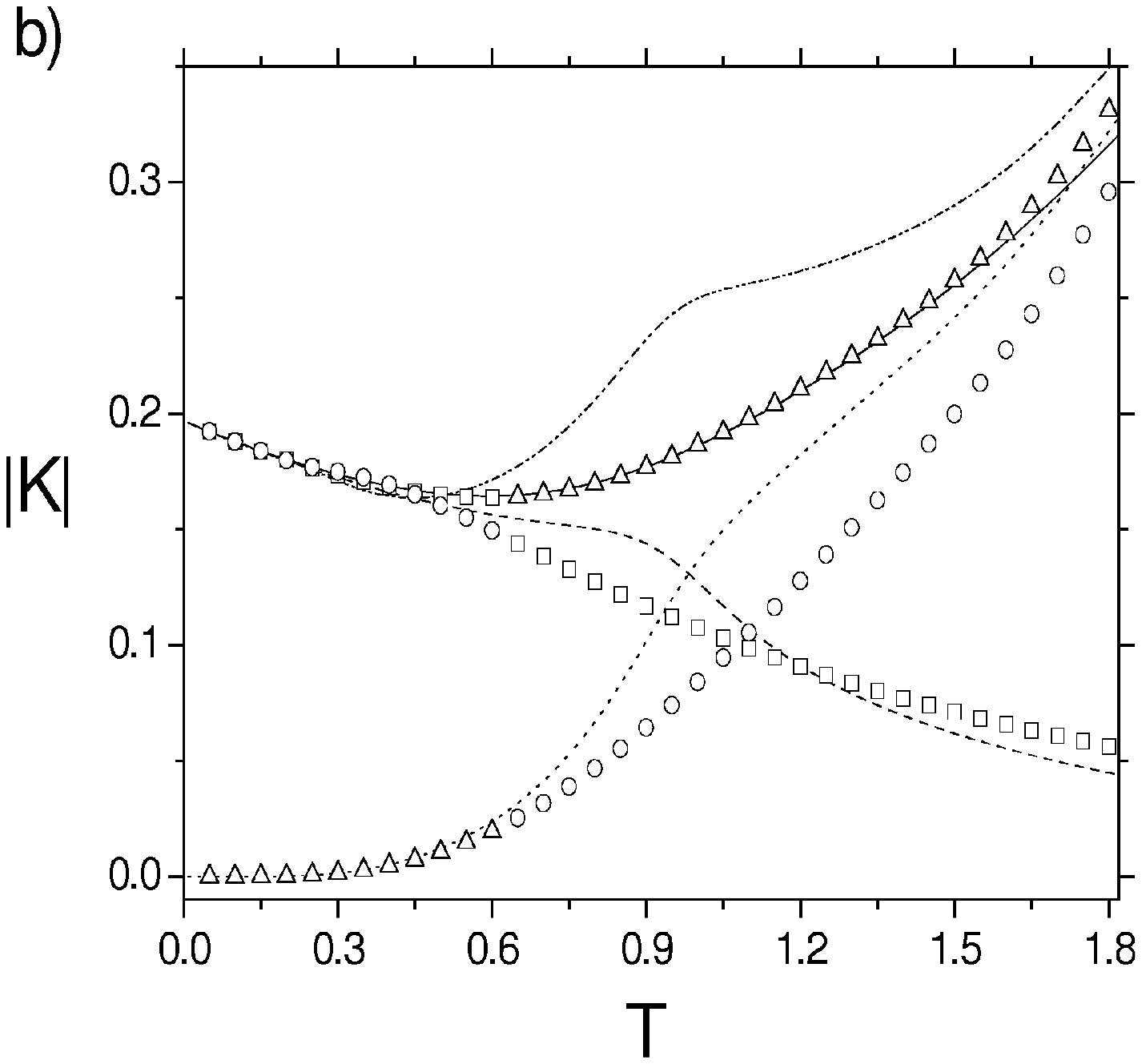}}
\vspace{0.2cm}
\centerline{\includegraphics[width=16pc]{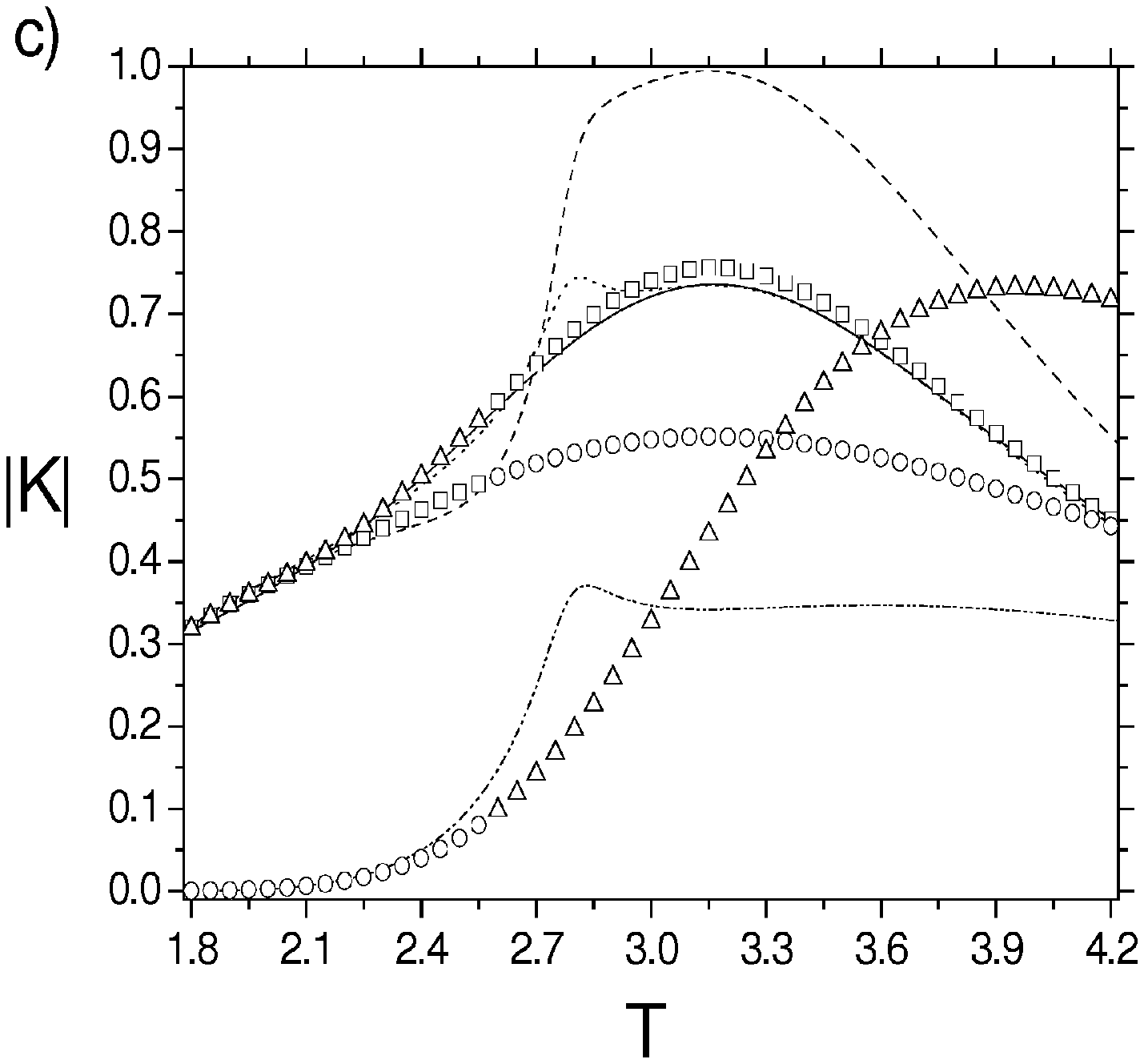}
\hspace{0.2cm}
\hspace{1cm}\includegraphics[width=16pc]{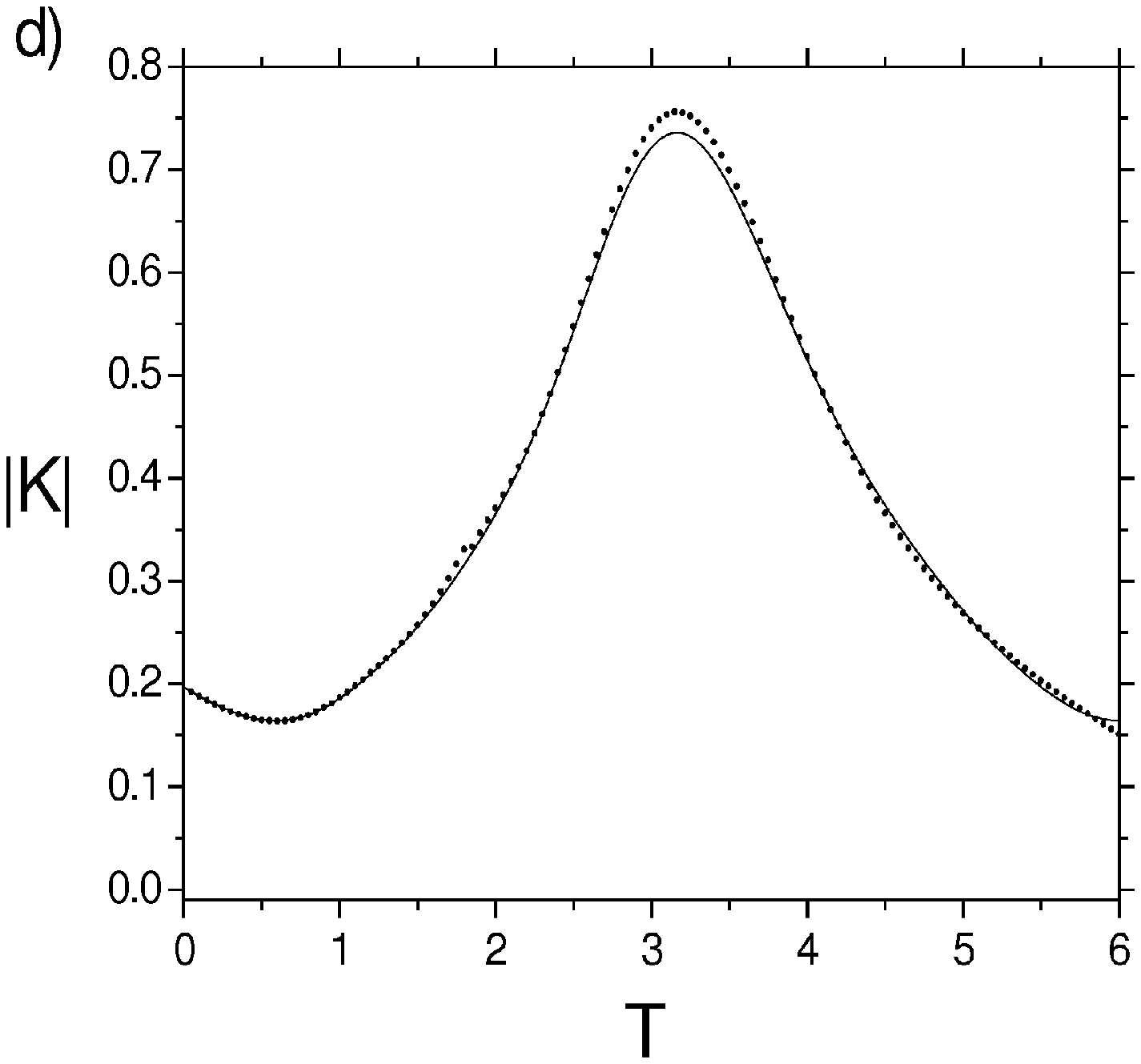}}
\caption{\label{fig2}
Exact propagator (full line in all panels) and semiclassical results for $K^{(2)}_{sc}(z''^*,z',T)$
and $K^{un}_{sc}(z''^*,z',T)$ for the quartic oscillator. Panel (a) shows the exact result and the
individual contribution of each family of complex trajectories to $K^{(2)}_{sc}$: $f1$ (dashed),
$f2$ (dotted) and $f3$ (dash-dotted). Panel (b) shows the propagator for $T<1.8$ and displays the
exact result, the contributions of $f1$ (dashed) and $f2$ (dotted) and the combined contributions
of $f1$ and $f2$ (dash-dotted). The uniform propagator $K^{un}_{sc}$ calculated with $f1$ and
$f2$ is also shown, evaluated for each possible path: $C_1$ (squares), $C_2$ (circles)
and $C_3$ (triangles). Panel (c) focuses on the interval $1.8 < T <4.2$ and shows the exact result,
the contributions of $f2$ (dotted) and $f3$ (dash-dotted) and the combined contributions of $f2$
and $f3$ (dashed). The uniform propagator calculated with $f2$ and $f3$ is also shown along the
three independent paths. Panel (d) shows the final result obtained with the uniform formula
(dots) and exact result.}
\end{figure}

In order to find complex trajectories satisfying the boundary
conditions (\ref{bbuv}) we follow the recipe of Rubin and
Klauder~\cite{Klau95} and define
\begin{equation}
Q(0)= q' + w \quad \mathrm{and} \quad P(0)= p' + i \frac{b}{c}w,
\label{inicon}
\end{equation}
where $w=\alpha+i\beta$ is a complex number. It is easy to check
that this choice of initial conditions satisfy the first of the
boundary conditions (\ref{bbuv}) for all values of $w$. The idea,
therefore, is to propagate trajectories for all possible $w$,
picking those satisfying the second of the boundary
conditions~(\ref{bbuv}). Specifically, for each $w$ we propagate
the complex trajectory starting from (\ref{inicon}) and calculate
\begin{equation}
v(T) \equiv v_T =
\frac{1}{\sqrt{2}}\left(\frac{Q(T)}{b}-i\frac{P(T)}{c}\right) \equiv \frac{1}{\sqrt{2}}
\left(\frac{Q''}{b}-i\frac{P''}{c}\right),
\end{equation}
where $Q(T)$ and $P(T)$ are complex and $Q''$ and $P''$ are real
variables, obtained by taking the real and imaginary parts of
$v_T$. The trajectories for which $(Q'',P'')=(q'',p'')$ are the
ones needed to calculate $K_{sc}^{(2)}(z''^*,z',T)$ and
$K_{sc}^{un}(z''^*,z',T)$.

Notice that the origin $\alpha=\beta=0$ corresponds to the real
trajectory starting from $Q(0)=q'$, $P(0)=p'$. Therefore, the
larger the $|w|$ the more complex is the corresponding trajectory
and the smaller its contribution to the propagator. Thus, we can
restrict our search to a small vicinity of the origin in the
complex $w$ plane.

By fixing $q'$, $p'$ and $T$, the values of the resulting $Q''$ and
$P''$ can be seen as function of $w$. In Fig.~\ref{fig1} we
represent the curves of constant $Q''$ superimposed with those of
constant $P''$ for different values of $T$ in the $(\alpha,\beta)$
plane. Stars identify the points where $Q''=q''=0.5$ and
$P''=p''=0.5$, our desired final values. As $T$ is changed, the
position of the stars move in the $w$ plane, forming families of
contributing trajectories. Circles (only in Fig.~\ref{fig1}(e)) are
the same as stars, but their contributions were not included
because they lie very far from the origin.

White stars in Fig.~\ref{fig1} represent trajectories belonging to
the family $f1$, which are close to the origin when $T=0$ but move
away as $T$ increases. Grey stars refer to the family $f2$.
Trajectories belonging to this family start off quite complex, but
approach the origin of the $w$ plane as $T$ tends to $3.2$,
approximately. After that, they move away from the origin again.
Finally, trajectories of the family $f3$ are represented by black
stars. They give important contributions for $T > 2$. Other
trajectories satisfying these same boundary conditions exist (see
Fig.~\ref{fig1}(e)) but their contribution to the propagator is not
significant.

For $T=0$, a contour plot like those in Fig.~\ref{fig1} would
display a grid of straight lines, vertical for $Q''=const.$ and
horizontal for $P''=const.$, with the only contributing trajectory
lying at $\alpha=1/4,~\beta=-5/4$. For short values of $T$, as in
Fig.~\ref{fig1}(a) for $T=0.06$, the grid deforms slightly and the
contributing trajectory moves away from the origin. For $T=0.24$,
Fig.~\ref{fig1}(b), there is still a single contributing
trajectory, but two defects on the grid are seen approaching the
origin. As discussed in Refs.~\cite{Klau95,Agu05}, these defects
are critical points of the map $v_T=v_T(w)$, where $\frac{\partial
Q''}{\partial w} = \frac{\partial P''}{\partial w}=0$.  In
addition, if the second derivatives of $Q''$ and $P''$ are
non-zero, it can be shown that the map becomes two-to-one in the
vicinity of the defects, implying two different trajectories
(corresponding to two distinct initial conditions) satisfying the
same boundary conditions~(\ref{bbuv}). These defects, therefore,
can be identified with the phase space caustics.

For $T=0.70$ and $T=1.02$, panels (c) and (d), two complex
trajectories can be seen and for $T=2.20$ and $T=2.70$, panels (e)
and (f), three of them are close to the origin. For $T=2.70$, the
trajectories belonging to $f2$ and $f3$ are the closest they get to
each other, moving away for larger $T$. The effect of the nearby
caustic should be pronounced at this point.

Figure~\ref{fig2}(a) shows the exact propagator and the individual
contribution of each family, $f1$, $f2$ and $f3$, to the propagator
$K^{(2)}_{sc}(z''^*,z',T)$ for $0 \leq T \leq 6$. For very short
times the contribution of $f1$ is clearly dominant, and reproduces
very well the exact results by itself. For $T > 0.5$ the family
$f2$ becomes more and more important, whereas the contribution of
$f1$ decreases. However, for $T\approx1.0$ the exact curve is not
reproduced by $f1$ nor by $f2$, suggesting that both families
should be included simultaneously.

Figure~\ref{fig2}(b) shows a detailed plot for the interval $0 \leq
T<1.8$ of such a combined contribution. The result obtained is
clearly not accurate. However, by looking at Fig.~\ref{fig1}, we
notice that these trajectories are close to a phase space caustic,
and, therefore, a divergent behavior is expected. Results for
$K^{un}_{sc}(z''^*,z',T)$ are presented in the same panel (b) for
each possible integration path, $C_1$, $C_2$ and $C_3$ (see
equation (\ref{int7})). Choosing the path $C_1$ for $0 \leq
T<0.65$, and $C_3$ for $0.65 \leq T <1.8$, the exact result is
satisfactorily reproduced.

Returning to Fig.~\ref{fig2}(a), we noticed that, after
$T\approx1.8$, the family $f2$ reproduces the exact result by
itself, except for the vicinity of $T\approx2.7$. As the family
$f3$ gives relevant contributions in this interval, we plot the
combined contributions of $f2$ and $f3$ in Fig.~\ref{fig2}(c). The
resulting curve is clearly inaccurate and the reason is, once
again, the proximity of these trajectories to a caustic, as can be
seen from  Fig.~\ref{fig1}(f). Thus, we use again the uniform
formula $K^{un}_{sc}(z''^*,z',T)$ (now for families $f2$ and $f3$),
and the resulting curves are shown in Fig.~\ref{fig2}(c). Path
$C_3$ should be chosen for $1.8 \leq T<2.55$, and $C_1$, for $T
\geq 2.55$, so that the exact curve is well reproduced.

Fig.~\ref{fig2}(d) shows the uniform approximation for the whole
time interval $0<T<6$, obtained by joining four pieces of the
propagator: (a) families $f1$ and $f2$ with path $C_1$ for
$T<0.65$; (b) families $f1$ and $f2$ with path $C_3$ for $0.65 \leq
T<1.8$; (c) families $f2$ and $f3$ with path $C_3$ for $1.8 \leq
T<2.55$; (d) families $f2$ and $f3$ with path $C_1$ for $T \geq
2.55$.

We note that the trajectories of the $f3$ family do not seem to
contribute to Eq.~(\ref{sp}), since their inclusion makes the
results worse everywhere. However, as they become partners of the
contributing trajectories belonging to $f2$ (when these approach
the caustic), their contributions become fundamental to control the
divergence of the family $f2$ when Eq.~(\ref{uform}) is used.

\section{The Nelson potential}
\label{nelson}

\begin{figure}
\includegraphics[width=16pc]{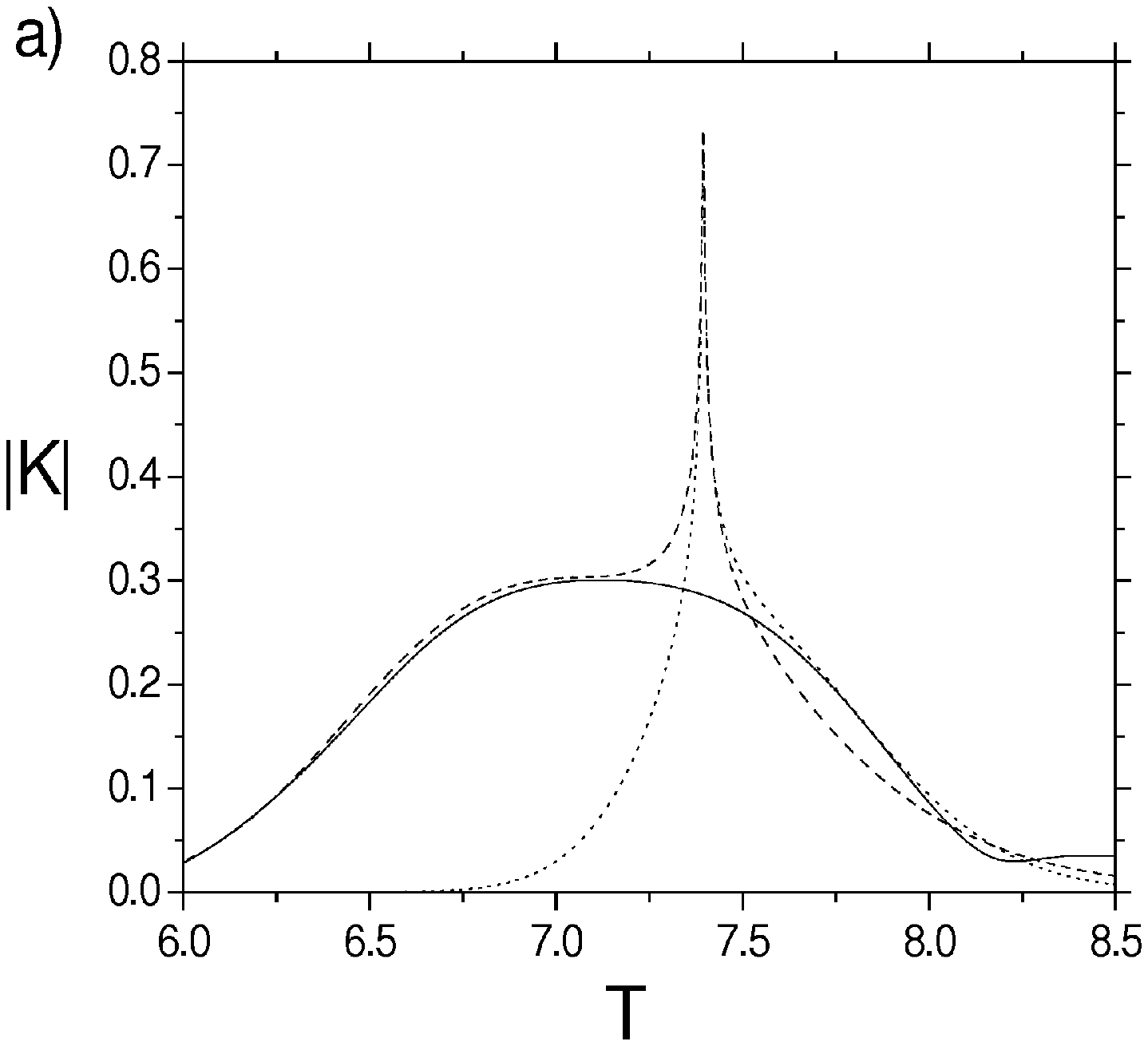}\hspace{1cm}
\includegraphics[width=16pc]{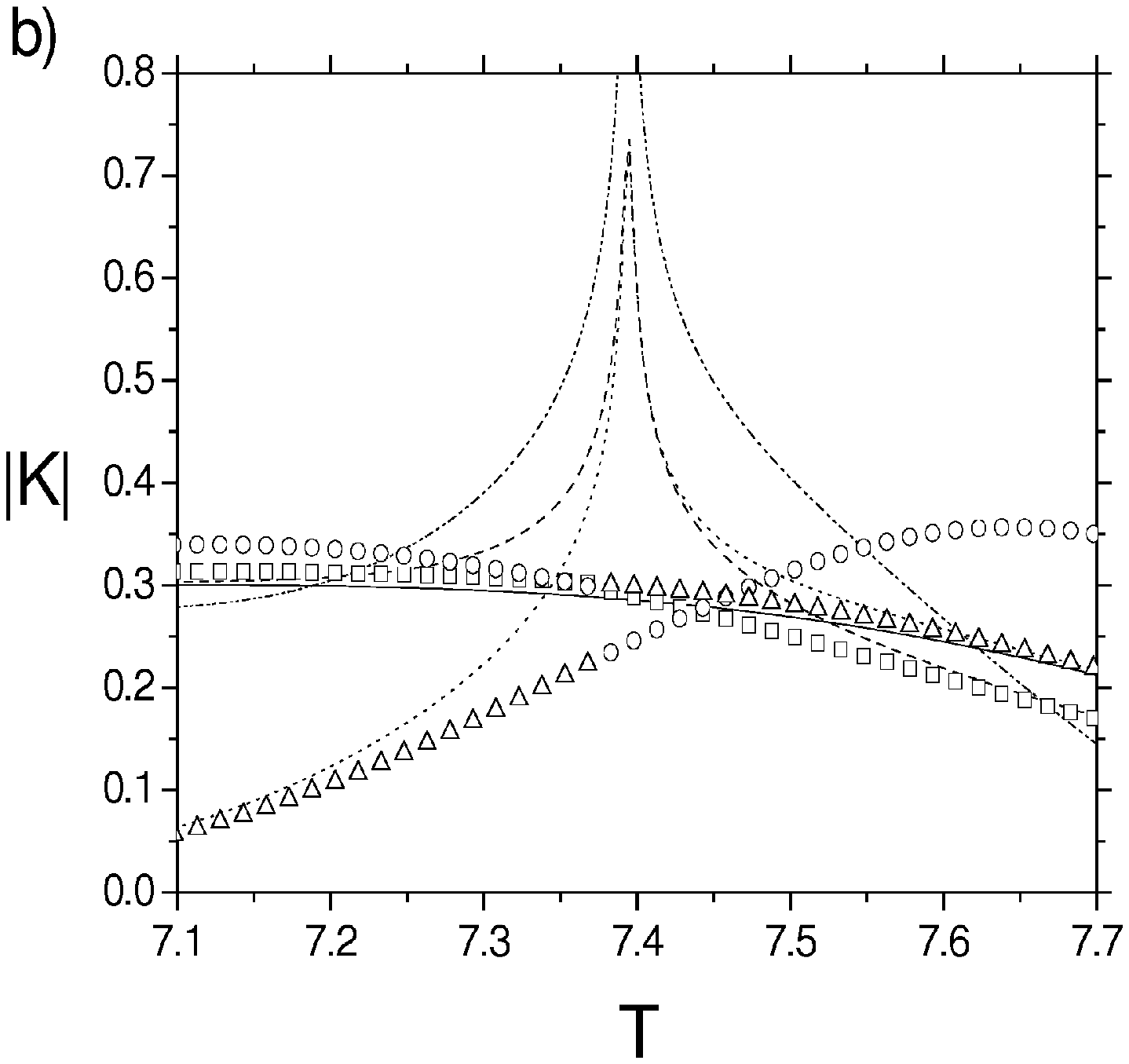}\\
\includegraphics[width=16pc]{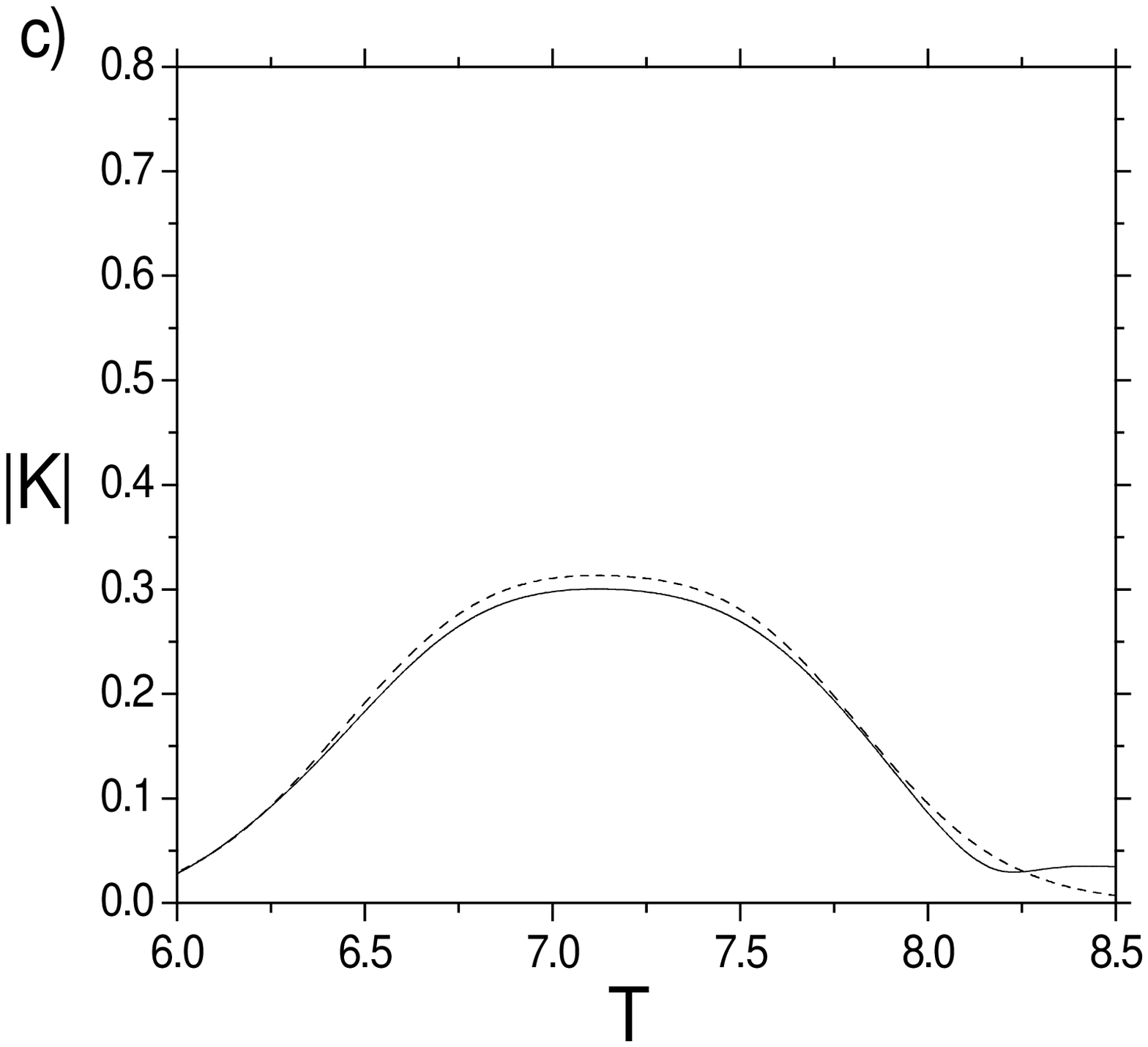}\hspace{1cm}
\begin{minipage}[b]{19pc}
\caption{\label{fig3}
Comparison between the exact propagator and the two semiclassical
approaches for the Nelson potential. Panel (a) shows the exact curve
(full line) and the individual contribution of two families of complex
trajectories (dashed lines for $f1$ and dotted lines for $f2$) to $K^{(2)}_{sc}$. Panel (b)
shows a blow up in the caustic region. The dash-dotted line corresponds
to the combined contribution of the two families of trajectories. The uniform
propagator is also shown for the paths $C_1$ (squares), $C_2$ (circles)
and $C_3$ (triangles). Panel (c) shows the exact (full line) and final semiclassical
 results (dashed line).}
\end{minipage}
\end{figure}

In the previous sections, both the theory and the numerical example
were presented for one-dimensional systems only. Here we show an
application for a two-dimensional chaotic system. The derivation of
expressions equivalent to~(\ref{sp}) and (\ref{uform}) for this
case can be found in references~\cite{ribeiro1,rib07a,garg07}. The
second order and uniform semiclassical formulas are given by
\begin{equation}
K_{sc}^{(2)}(\mathbf z''^*,\mathbf z',T)=   \sum_{\mathrm{traj.}}
   \sqrt{ \frac{1}{\left| \det \mathbf{\mathrm{M}_{vv}} \right|}} \,
   \exp{ \left\{ \frac{i}{\hbar} \, (\mathcal{S}+\mathcal{G})
   -\frac{1}{2}\left(|\mathbf z'|^2+|\mathbf z''|^2\right)
   \right\}} \label{sp2}
\end{equation}
and
\begin{equation}
K_{sc}^{un}(\mathbf z''^*,\mathbf z',T)=
i\sqrt\pi~e^{\mathcal A}
\left\{
\left(\frac{h_2-h_1}{\sqrt \mathcal B}\right)\mathrm{f'_i}(\mathcal B)+
(h_1+h_2)\mathrm{f_i}(\mathcal B)
\right\} \label{uform2}
\end{equation}
where $|\mathbf{z} \rangle \equiv |z_x \rangle \otimes |z_y
\rangle$  is the direct product of two 1-D states,
\begin{eqnarray}
\mathcal{S}({\mathbf z''}^*,\mathbf z', T)&=&\int_{0}^{T}
\left[\frac{i\hbar}{2}\left(\dot{\mathbf u}~\mathbf v - \mathbf u~\dot{\mathbf v}\right)-\tilde{H}\right] dt
-\frac{i\hbar}{2}\left[{  \mathbf u(T)~ {\mathbf z''}^* + \mathbf z'~\mathbf v(0) } \right],\label{action2d}\\
\mathcal{G}({\mathbf z''}^*,\mathbf z', T) &=&\frac{1}{2} \int_{0}^{T}
\left[\frac{\partial^{2}\tilde{H}}{\partial u_x \; \partial v_x} +
\frac{\partial^{2}\tilde{H}}{\partial u_y \; \partial v_y} \right]\, dt ,
\label{gfactor2}
\end{eqnarray}
\begin{eqnarray}
\left(
   \begin{array}{c}
   \delta \mathbf{u}''\\
   \delta \mathbf{v}''\\
   \end{array}
\right) = \left(
   \begin{array}{cc}
   \mathbf{\mathrm{M}_{uu}}   & \mathbf{\mathrm{M}_{uv}}   \\
   \mathbf{\mathrm{M}_{vu}}   & \mathbf{\mathrm{M}_{vv}}   \\
   \end{array}
\right) \left(
   \begin{array}{c}
   \delta \mathbf{u}'\\
   \delta \mathbf{v}'\\
   \end{array}
\right) \label{eq5}
\end{eqnarray}
and
\begin{equation}
h_{1,2}=
\sqrt{\mp  \frac{\sqrt{\mathcal B}}{\left. \left(
\det \mathrm{M_\mathbf{vv}}\right)\right|_{\mathbf u''_{1,2}}} }
~e^{\frac{i}{\hbar}{\left.\mathcal G\right|_{z_{1,2}}}}.
\label{g1g2}
\end{equation}

For the numerical application we have chosen the {\em Nelson Hamiltonian},
\begin{displaymath}
H = \frac{1}{2}(p_x^2+p_y^2)+{\left( y - x^2/2 \right)}^2 + 0.05 x^2.
\end{displaymath}
This system has been widely studied both classically~\cite{michel}
and quantum mechanically~\cite{ribeiro1}. In particular, the
coherent state propagator and its semiclassical quadratic
approximation were investigated in great detail in~\cite{ribeiro1}
and some strongly divergent semiclassical behavior due to the
presence of phase space caustics were identified. Here we revisit
this problem, using the same parameters for which the caustics were
found and apply the uniform formula.

The widths of $|\mathbf{z}'\rangle$ and $|\mathbf{z}''\rangle$ were
fixed to $b_x = b_y = 0.2$ and $\hbar = 0.05$. The eight remaining
parameters fixing the initial and final coherent states are:
$x'=x''=0.72$, $y'=y''=0.24$, $p_x'=p_x''=-0.75$ and
$p_y'=p_y''=-0.63$, which refer to a diagonal element of the
propagator.

Two families of trajectories ($f1$ and $f2$) contribute to the
propagator~(\ref{sp2}) when $T$ is varied from 6 to 8.5, as shown
in Fig.~\ref{fig3}(a). Family $f1$ reproduces very well the exact
propagator for the range $6<T<7.2$, while $f2$ does the same for
$7.6<T<8.5$. In the vicinity of $T\approx7.4$, the semiclassical
result shows a divergent behavior. As shown in Fig.~\ref{fig3}(b),
the combined contributions of these families only makes things
worse. A careful analysis of the classical phase space shows that
this region is close to a caustic~\cite{ribeiro1}. Therefore, it is
indicated to use expression~(\ref{uform2}) instead of~(\ref{sp2}).
The results for the uniform approximation are shown in
Fig.~\ref{fig3}(b) for the paths $C_1$, $C_2$ and $C_3$. Taking
path $C_1$ for $T<7.38$ and $C_3$ for $T \geq 7.38$ kills the
divergence and the exact result is reproduced, as shown in
Fig.~\ref{fig3}(c).

\section{Final Remarks}
\label{final}

Focal points and caustics are well known sources of inaccuracies in
semiclassical approximations. The most systematic way to derive
semiclassical formulas that avoid such divergences is to use the
method of Maslov, which consists basically of two steps. The first
step is to compute the semiclassical propagator in a dual
representation. The caustics in the original and the dual
representations lie generically in different regions of phase
space, so that at least one of the propagators is well behaved at
any given point. The second step is to transform back to the
representation of interest doing the corresponding integral by the
stationary phase approximation, but expanding beyond the second
order.

Coherent states lack a natural dual representation and we have
defined one such re\-pre\-sen\-ta\-tion in \cite{prl}. Although the
transformation leading from the Bargmann to the dual form is not a
simple Fourier transform like in the case of position and momentum,
it leads to a similar Legendre transformation of the action in the
semiclassical limit. With this representation in hand the second
step of the Maslov method can be carried out as usual. It is not
clear at this point if this alternative representation can be
useful in other contexts and work in this direction is in progress.

Finally we remark that one of the few semiclassical approaches that
naturally avoids caustics is the initial value representation of
Herman and Kluk \cite{hk,miller,kay} (see also \cite{pollak}), for
which the tangent matrix elements, that go to zero at the caustic,
appear in the numerator and do not lead to divergencies. However,
convergence problems due to highly oscillatory contributions have
been reported for chaotic systems \cite{hkp}, which also required
the development of additional techniques and methods.

\ack
The authors acknowledge financial support from CNPq, FAPESP
and FINEP. ADR especially acknowledges FAPESP for the fellowship
$\#$ 04/04614-4. {\it Wilhelm und Else Hereaus Foundation} and {\it Instituto do
Mil\^enio de Informa\c{c}\~ao qu\^antica -- CNPq} are gratefully
acknowledged for providing financial support for the Blaubeuren
meeting. We also acknowledge AFR de Toledo Piza, Marcel
Novaes and Fernando Parisio for their important contributions.


\end{document}